\documentstyle[prl,aps,twocolumn,psfig]{revtex}

\begin{document}

\twocolumn[\hsize\textwidth\columnwidth\hsize\csname @twocolumnfalse\endcsname
\title{Jet Quenching in the Opposite Direction of a Tagged Photon
  in High-Energy Heavy-Ion Collisions}
\author{Xin-Nian Wang}
\address{Nuclear Science Division, MS 70A-3307, \\
\baselineskip=12pt Lawrence Berkeley Laboratory, Berkeley, CA 94720}
\author{Zheng Huang and Ina Sarcevic}
\address{Department of Physics, University of Arizona,
Tucson, AZ 85721}

\date{February 16, 1996}
\maketitle

\begin{abstract}

We point out that events associated with large $E_T$ direct photons 
in high-energy heavy-ion collisions can be used to study jet energy 
loss in dense matter. In such events, the $p_T$ spectrum of 
charged hadrons from jet fragmentation in the opposite direction 
of the tagged photon is estimated to be well above the background
which can be reliably subtracted at moderately large $p_T$. 
We demonstrate that comparison between the extracted fragmentation 
function in $AA$ and $pp$ collisions can be used to determine the 
jet energy loss and the interaction mean-free-path in the dense matter 
produced in high-energy heavy-ion collisions.
\end{abstract}
]


Large transverse momentum jets, among many other hard processes
in high-energy heavy-ion collisions, have been proposed as
effective probes of the transient dense matter.
{}For example, an enhanced acoplanarity and 
energy imbalance of two back-to-back jets can be used to study 
multiple scatterings of a parton inside a dense medium \cite{aco}. 
Study of large $p_T$ jets can also probe their energy loss 
due to inelastic scatterings inside a dense matter or a quark-gluon
plasma \cite{qn1}.
Because of the enormous background in high-energy heavy-ion 
collisions,
the conventional calorimetric study cannot measure the jet
energy to such an accuracy as required to determine the energy loss.
Alternately, single-particle inclusive $p_T$-spectrum has been
shown to be sensitive to the jet energy loss \cite{qn2}. Since
the single-particle spectrum is a convolution of the jet
production cross section and the jet fragmentation functions,
the suppression of produced hadrons at a fixed $p_T$ results from
jet quenching with a wide range of initial transverse
energies, thus making it difficult to measure directly the
modification of jet fragmentation for a given transverse
energy.

In this Letter, we propose to study the jet quenching in high-energy 
heavy-ion collisions by measuring the $p_T$ distribution of
charged hadrons in the opposite direction of a tagged direct
photon. A direct photon is produced by quark-antiquark annihilation
or quark(antiquark)-gluon Compton scatterings in which
a gluon or quark(antiquark) jet is also produced in the
opposite direction of the photon. By tagging a direct photon
with a given transverse energy $E_T^{\gamma}$, one can avoid
the uncertainties associated with the jet production cross 
section. One can also determine the initial transverse energy 
of the produced jet, $E_T\approx E_T^{\gamma}$,
from momentum conservation, modulo calculable corrections
from initial state radiations. At collider energies and sufficiently
large $E_T^{\gamma}$, the Cronin effect due to
multiple scatterings during the initial interaction stage
is also negligible \cite{review}.
We shall use perturbative QCD to show that the $p_T$ spectrum of charged 
hadrons with moderate $p_T$ in the backward direction of a 
direct photon is a very good approximation of the jet 
fragmentation function which can thus be reliably extracted.
We shall also study the sensitivity of the modification
of the jet fragmentation functions in heavy-ion collisions
to the energy loss of jets and the jet interaction
mean-free-path inside a dense matter.

Let us consider events with a direct photon in the central
rapidity region, $|y|\leq \Delta y/2$, $\Delta y=1$. 
{}For sufficiently large $E_T^{\gamma}$ of the photon, the rapidity 
distribution of the associated jet is also centered around 
zero rapidity with a comparable width. If the azimuthal 
angle of the photon is $\phi_{\gamma}$ and 
$\bar{\phi}_{\gamma}=\phi_{\gamma}+\pi$, most of the hadrons
from the jet fragmentation will fall into the kinematic region,
$(|y|\leq \Delta y/2, |\phi-\bar{\phi}_{\gamma}|\leq \Delta\phi/2)$,
where one can take $\Delta\phi=2$ according to the jet profile
as measured in high-energy $p\bar{p}$ collisions \cite{ua1}.
Given the jet fragmentation functions $D_{h/a}(z)$, with $z$ 
the fractional momenta of the hadrons, one can calculate the 
differential $p_T$ distribution of hadrons from the jet 
fragmentation in the kinematical region $(\Delta y,\Delta \phi)$,
\begin{equation}
  \frac{dN_{ch}^{jet}}{dyd^2p_T}=
  \sum_{a,h}r_a(E_T^{\gamma})\frac{D_{h/a}(p_T/E_T)}{p_T E_T} 
  \frac{C(\Delta y,\Delta\phi)}{\Delta y\Delta\phi}, \label{eq:frg1}
\end{equation}
where $C(\Delta y,\Delta\phi)=\int_{|y|\leq \Delta y/2}dy
\int_{|\phi-\bar{\phi}_{\gamma}|
\leq \Delta\phi/2}d\phi f(y,\phi-\bar{\phi}_{\gamma})$
is an overall factor and $f(y,\phi)$ is the hadron 
profile around the jet axis. 
The summation is over both jet ($a$) and hadron species ($h$),
and $r_a(E_T^{\gamma})$ is the fractional production cross section of
$a$-type jet associated with the direct photon.
We define $D^{\gamma}(z)=\sum_{ah}r_a(E_T^{\gamma})D_{h/a}(z)$ 
as the inclusive fragmentation function.
$C(\Delta y,\Delta\phi)$ is the acceptance factor for finding 
the jet fragments in the given kinematic range.
We find $C(\Delta y,\Delta\phi)\approx 0.5$
at $\sqrt{s}=200$ GeV, independent of the photon energy $E_T^{\gamma}$,
using HIJING \cite{hijing} Monte Carlo simulations for 
the given kinematic cuts.
{}For a fixed $E_T^{\gamma}$, the jet $E_T$ has a smearing around
$E_T^{\gamma}$ caused by initial state radiations. One should
therefore average Eq.~(\ref{eq:frg1}) over such a smearing.
The resultant spectrum is very well approximated by
Eq.~(\ref{eq:frg1}) with $E_T=E_T^{\gamma}$ \cite{hw2},
as will be shown by comparison with explicit HIJING Monte
Carlo simulations.

To calculate the background for the  photon-tagged jet
fragmentation from particle production in a
normal central nucleus-nucleus collision, 
one convolutes the fragmentation functions with the jet 
cross sections \cite{owens},
\begin{equation}
  \frac{dN_{ch}^{AA}}{dyd^2p_T}=K\int d^2r\sum_{abcdh}
  \int_{x_{amin}}^1 dx_a \int_{x_{bmin}}^1 dx_b 
  f_{a/A}(x_a,r)f_{b/A}(x_b,r)\frac{D_{h/c}(z_c)}{\pi z_c}
  \frac{d\sigma}{d\hat{t}}(ab\rightarrow cd), \label{eq:nch}
\end{equation}
where $z_c=x_T(e^y/x_a +e^{-y}/x_b)/2$, 
$x_{bmin}=x_ax_Te^{-y}/(2x_a-x_Te^y)$,
$x_{amin}=x_Te^y/(2-x_Te^{-y})$, and $x_T=2p_T/\sqrt{s}$.
The $K\approx 2$ factor accounts for higher order corrections \cite{xwke}.
The parton distribution density in a nucleus,
$f_{a/A}(x,r)=t_A(r)S_{a/A}(x,r)f_{a/N}(x)$, is assumed to
be factorizable into the nuclear thickness function 
$t_A(r)$ (with normalization $\int d^2r t_A(r)=A$),
parton distribution in a nucleon $f_{a/N}(x)$ and the
parton shadowing factor $S_{a/A}(x,r)$ which we take the
parametrization used in HIJING model \cite{hijing}. In our notation,
the scale dependences of the parton distributions $f_{a/N}(x,Q^2)$ 
and the fragmentation functions $D_{h/a}(z,Q^2)$ are 
implicit, which we take to be $Q=E^{\gamma}_T$.

Jet fragmentation functions have been studied extensively
in $p\bar{p}$, $ep$ and $e^+e^-$ experiments \cite{mattig}. We
will use the parametrizations of both $z$ and $Q^2$ dependence 
of the most recent analysis \cite{bkk} for the unmodified 
fragmentation functions $D^0_{h/a}(z)$, in which only pions 
and kaons are included.  We will use 
the MRS D$-'$ parametrization of the parton distributions \cite{mrs}. 
The resultant single-particle $p_T$
spectra from Eq.~(\ref{eq:nch}) for $pp$ and $p\bar{p}$
collisions at different energies agree well with the experimental
data at moderate $p_T\geq 2$ GeV/$c$ \cite{hw2} where particle 
production from soft processes is expected to be small.
Shown in Fig.~1 are the differential $p_T$ distributions from
the fragmentation of a photon-tagged jet with $E_T^{\gamma}=$15, 20 GeV
and the underlying background of normal central $Au+Au$
collisions at $\sqrt{s}=200$ GeV.
The points are HIJING simulations of 10K events and solid
lines are numerical results of Eqs.~(\ref{eq:frg1}) and
(\ref{eq:nch}), in both cases no medium effects have been
considered in the fragmentation functions. The background 
in $pp$ collisions is about 1200 times smaller than $Au+Au$.

In heavy-ion collisions, produced partons will experience
secondary scatterings and induced radiation which will
drive the system toward equilibrium. As a result,
large momentum partons will lose part of their energy
before they escape and fragment into hadrons. There have
been many studies on the energy loss of a propagating
parton inside a medium. It is believed that radiative energy
loss dominates even when the Landau-Pomeranchuk-Migdal suppression 
is taken into account \cite{lpm1,lpm2}. 
While a dynamical study of the jet propagation
and the modification of the hadronization is more desirable, 
we will use a phenomenological model here to demonstrate how
sensitive our proposed measurement in the photon-tagged
events to the interactions and the average energy loss 
suffered by a parton in a dense medium.

We restrict ourselves to the central rapidity region so that
a parton will only propagate in the transverse 

\vspace{0.8in}

\noindent  
direction in a cylindrical system. The parton will not hadronize
inside a deconfined quark-gluon plasma. In a hadronic
medium, we assume that the fragmentation functions can be 
approximated by their forms in vacuum. We only study the
effects of radiative energy loss.
Given the inelastic scattering mean-free-path, $\lambda_a$, 
the probability for a parton to scatter $n$ times
within a distance $\Delta L$ before it escapes the system 
is assumed to be
\begin{equation}
  P_a(n) = \frac{(\Delta L/\lambda_a)^n}{n!} e^{-\Delta L/\lambda_a}.
\end{equation}
If we assume the average energy loss per scattering suffered by 
the parton is $\epsilon_a$, the modified fragmentation 
functions can be approximated as,
\begin{eqnarray}
  D_{h/a}(z,\Delta L,Q^2)& =&
  \frac{1}{C^a_N}\sum_{n=0}^NP_a(n)\frac{z^a_n}{z}D^0_{h/a}(z^a_n,Q^2)
  \nonumber \\
  &+&\langle n_a\rangle\frac{z'_a}{z}D^0_{h/g}(z'_a,Q_0^2), 
  \label{eq:frg2}
\end{eqnarray}
where $z^a_n=z/(1-n\epsilon_a/E_T)$, $z'_a=zE_T/\epsilon_a$ and
$C^a_N=\sum_{n=0}^N P_a(n)$. We limit the number of inelastic
scatterings to $N=E_T/\epsilon_a$ by energy conservation.
{}For large values of $N$, the average number of scatterings
within a distance $\Delta L$ is approximately 
$\langle n_a\rangle \approx \Delta L/\lambda_a$.
The first term corresponds to the fragmentation of the 
leading partons with reduced energy $E_T-n\epsilon_a$
and the second term comes from the emitted gluons each
having energy $\epsilon_a$ on the average.
{}For simplification, we have neglected the
fluctuation in the energy carried by each emitted gluon
and its possible rescatterings. The inelastic scatterings suffered
by the leading parton are normally not hard. Therefore, 
we also assume the scales in the fragmentation functions 
of the emitted gluons are given by the initial value 
$Q^2_0=2.0 \; {\rm GeV}^2$. Since the emitted gluons will 
only contribute to hadrons with very small fractional energy, 
the final modified  fragmentation function in the moderately 
large $z$ region is not sensitive to the actual radiation spectrum
and the scale dependence of the fragmentation.

\vspace{0.4in}
\begin{figure}
\centerline{\psfig{figure=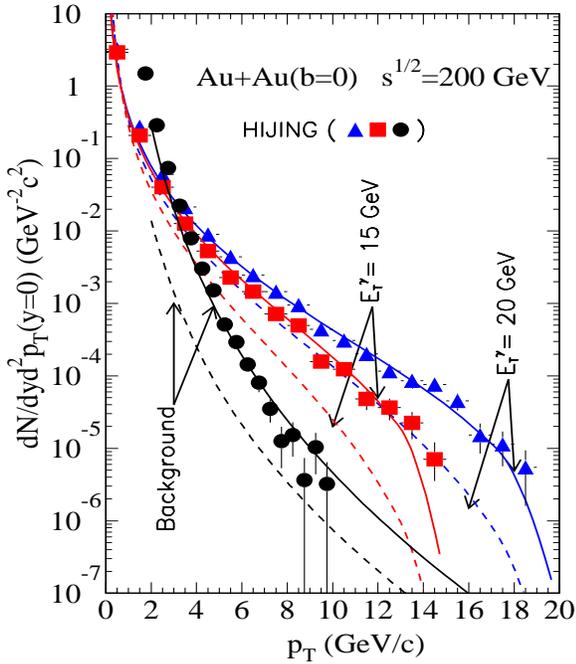,width=3in,height=3.5in}}
\caption{The differential $p_T$ spectrum of charged particles
  from the fragmentation of a photon-tagged jet with 
  $E_T^{\gamma}=$15, 20 GeV
  and the underlying background in central $Au+Au$ collisions at
  $\protect\sqrt{s}=200$ GeV. The direct photon is restricted to 
  $|y|\leq \Delta y/2=0.5$. Charged particles are limited to the
  same rapidity range and in the opposite direction of the photon,
  $|\phi-\phi_{\gamma}-\pi|\leq \Delta\phi/2=1.0$. Solid lines
  are perturbative calculations and points are HIJING
  simulations of 10K events. The dashed lines are calculations
  with jet energy loss, $dE_q/dx=1$ GeV/fm and the mean-free-path
  $\lambda_q=1$ fm.}
\end{figure}

Since the jet production rate is proportional to the number of
binary nucleon-nucleon collisions, the averaged inclusive
fragmentation function of a photon-tagged jet in a central
nucleus-nucleus collision is
\begin{equation}
  D^{\gamma}_{AA}(z)=\int \frac{d^2r t^2_A(r)}{T_{AA}(0)}\sum_{ah} 
  r_a(E_T^{\gamma})D_{h/a}(z,\Delta L) \; ,
\end{equation}
where $T_{AA}(0)=\int d^2r t^2_A(r)$ is the overlap function of
$AA$ collisions at zero impact-parameter. Neglecting the transverse
expansion, $\Delta L(r,\phi-\bar{\phi}_{\gamma})$ only
depends on the jet production position $(r,\phi)$.
Using Eq.~(\ref{eq:frg2}) in Eq.~(\ref{eq:nch}), 
we can calculate the single-particle
inclusive $p_T$ spectrum of normal central $AA$ collisions
taking into account jet quenching. 

In principle, $\epsilon_a$ and $\lambda_a$ are related 
to each other in a dynamical model \cite{lpm1,lpm2}. 
Phenomenologically, we can treat them as independent parameters.
Alternatively, we will vary $\lambda_a$ and 
$dE_a/dx=\epsilon_a/\lambda_a$ in our calculations.
The dashed lines in Fig.~1 are calculated with the modified 
fragmentation functions, with $dE_q/dx=1$ GeV/fm and $\lambda_q=1$ fm. 
We have assumed that the mean-free-path of a gluon is half and the 
energy loss is twice that of a quark. During the parton propagation, 
multiple scatterings can also change the direction of the parton
resulting in a sizable acoplanarity. Such an acoplanarity due 
to multiple scatterings is probably small as compared to 
that caused by initial state radiations for a large $E_T^{\gamma}$
photon. Thus, we assume the acceptance factor $C(\Delta y, \Delta\phi)$
to be the same as in $pp$ collisions.
One observes that there is significant
suppression of large $p_T$ particles both from the background
and jet fragmentation
in the opposite direction of a tagged photon due to jet quenching.
Since the number of particles at large $p_T\geq 4$ GeV/$c$ from 
the underlying background is substantially smaller than from the 
tagged jet fragmentation with and without jet quenching, one can accurately 
measure the jet fragmentation function from the $p_T$ distribution 
of charged particles in the opposite direction of the tagged photons,
given enough number of events. Once the background is subtracted,
one can push the limit to even smaller
$p_T\geq 2$ GeV/$c$, which corresponds to $z\sim 0.1$. 
One can then compare the fragmentation functions measured 
in $pp$, $pA$ or peripheral
$AA$ with central $AA$ collisions to obtain 
the modification due to jet quenching. 

\begin{figure}
\centerline{\psfig{figure=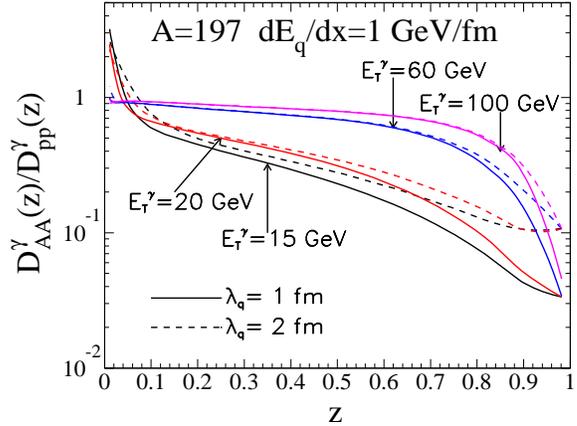,width=3in,height=2.25in}}

\caption{Ratio of the inclusive fragmentation function of
  a photon-tagged jet with and without energy loss in central
  $Au+Au$ collisions for a fixed $dE_q/dx=1$ GeV/fm.}
\end{figure}

To study the sensitivity of the modified inclusive fragmentation 
function to the energy loss, and the interaction 
mean-free-path, $\lambda$, we plot in Fig.~2 the ratio of the 
fragmentation functions with and without energy loss
for central $Au+Au$ collisions. There is enhancement of
soft particle production due to induced emissions, but only
at very small values of $z$. The fragmentation function
is suppressed for large range of $z$ due to energy loss.
{}For fixed $dE_q/dx=1$ GeV/fm, the suppression is delayed
to larger values of $z$ for larger jet energies. The most
optimal situation is when the average total energy loss
$\langle \Delta E_T\rangle$ is comparable to the initial
jet energy so that substantial suppression happens at
moderate values of $z$. From Figs.~2 and 1, we can see
that there is such a window of opportunity 
between $E_T^{\gamma}=10$ and 20 GeV at $\sqrt{s}=200$ GeV 
where the background is small.

{}For large values of $z>0.9$, particles from the leading
jets, which have suffered at least one inelastic scattering,
are completely suppressed. The remaining contribution
comes from only those jets that escape the system without
a single scattering.
The suppression factor is given by
$\langle\exp(-\Delta L/\lambda_a)\rangle$, independent
of jet energy $E_T$ and the energy loss $dE_a/dx$.
Therefore, one can determine the jet interaction
mean-free-path by measuring the suppression factor
of the jet fragmentation function at large $z>0.9$.
{}For intermediate values of $z\sim 0.2$--0.5, particles from
the leading partons  with reduced energy dominate
as far as $\epsilon_a \ll E_T$, a situation we will refer to 
as the ``soft emission'' scenario. Since the average 
total energy loss by the leading parton is 
$\langle \Delta E_{Ta}\rangle=\langle n_a \rangle\epsilon_a
=\langle \Delta L\rangle dE_a/dx$, the suppression factor 
should scale with $dE_a/dx$, depending very weakly on 
the mean-free-path.  Shown in Fig.~3 are the
suppression factor at $z=0.3$ as a function of $dE_q/dx$
for three different values of the mean-free-path. We see
that for the soft emission scenario, the suppression factor
scales and decreases almost linearly with $dE_q/dx$. At large
values of $dE_q/dx$ and $\lambda_q$, the average total
energy loss becomes comparable or equal to the initial jet 
energy $E_T$. In this ``hard emission'' scenario, particle
production from the emitted gluons and contributions from 
those jet partons which escape the system without any induced 
radiation become important. 
This is why the suppression factor saturates at larger values of $dE_q/dx$,
especially for large values of $\lambda_q$. Since the mean-free-path 
can be determined from the measured suppression factor
at large $z>0.9$ which is independent of $dE_q/dx$, additional
measurements of the suppression at intermediate $z=0.2$--0.4
will enable one to extract the energy loss.

\begin{figure}
\centerline{\psfig{figure=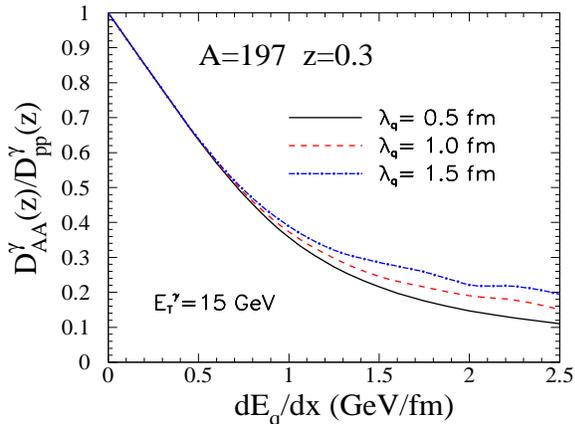,width=3in,height=2.25in}}

\caption{Ratio of the inclusive fragmentation function of
  a photon-tagged jet with and without energy loss in central
  $Au+Au$ collisions at $z=0.3$ as a function of $dE_q/dx$.}
\end{figure}

In summary, we have proposed to study jet energy loss
in high-energy heavy-ion collisions by measuring the inclusive
jet fragmentation function which can be extracted from the
differential $p_T$ spectrum of charged particles in the
opposite direction of a tagged direct photon. The background
to the jet fragmentation is estimated to be small for moderately
large $p_T$. We have also demonstrated that modification
of the jet fragmentation function due to jet quenching can
be used to obtain the energy loss  and the mean-free-path
of jet interaction inside the dense matter produced in
high-energy heavy-ion collisions.

We have not specified the energy dependence of the energy loss
in our calculation. In addition, the energy loss, $dE/dx$, might 
also depend on the distance that jet partons have traveled as
indicated by a recent study \cite{lpm2}. These dependences
can be studied experimentally by  varying the energy of the 
tagged photons in the collisions of different nuclei. These are
the subjects of further investigations \cite{hw2}.

We would like to thank J.~B.~Carroll, M. Gyulassy,
J.~W.~Harris and R. Thews for helpful discussions. 
This work was supported by the U.S. Department of Energy under Contract
Nos. DE-AC03-76SF00098, DE-FG03-93ER40792.
X.N.W was also supported by the
U.S. - Hungary Science and Technology Joint Fund J.F.No.378.


\begin{references}
\bibitem{aco}D. A. Appel, Phys. Rev. D {\bf 33}, 717 (1986);
J. P. Blaizot and L. D. McLerran, Phys. Rev. D {\bf 34}, 2739 (1986);
M. Rammerstorfer and U. Heinz, Phys. Rev. D {\bf 41}, 306 (1990);
S. Gupta, Phys. Lett. {\bf B347}, 381 (1995).
\bibitem{qn1}M. Gyulassy and M. Pl\"umer, Phys. Lett. {\bf B243}, 432 (1990);
M. Pl\"umer, M. Gyulassy and X.-N. Wang, Nucl. Phys. A 590, 511c (1995).
\bibitem{qn2}X.-N. Wang and M. Gyulassy, Phys. Rev. Lett. 
{\bf 68}, 1480 (1992).
\bibitem{review}X.-N. Wang, in {\it Quark-Gluon Plasma II}, R. C. Hwa (ed.)
(World Scientific, 1995).
\bibitem{ua1}UA1 Collab., G. Arnison {\it et al.}, Phys. Lett. {\bf B 172},
461 (1986); C. Albajar {et al.}, Nucl. Phys. {\bf B309}, 405 (1988).
\bibitem{hijing}X.-N. Wang and M. Gyulassy, Phys. Rev. D {\bf 44}, 3501 (1991);
Comp. Phys. Comm. {\bf 83}, 307 (1994).
\bibitem{hw2}Z. Huang and X.-N. Wang, to be published.
\bibitem{owens}J. F. Owens, Rev. Mod. Phys. {\bf 59}, 465 (1987).
\bibitem{xwke}K. J. Eskola and X.-N. Wang, Int. J. Mod. Phys. A {\bf 10}, 
3071 (1995).
\bibitem{mattig}P. M\"{a}ttig, Phys. Rep. {\bf 177}, 141 (1989).
\bibitem{bkk}J. Binnewies, B. A. Kniehl and G. Kramer, DESY-94-124,
hep-ph/9407347.
\bibitem{mrs}A. D. Martin, W. J. Stirling and R. G. Roberts,
                        Phys. Lett. {\bf B306}, 145 (1993).
\bibitem{lpm1}M. Gyulassy and X.-N. Wang, Nucl. Phys. {\bf B420}, 583 (1994);
X.-N. Wang, M. Gyulassy and M. Pl\"umer, Phys. Rev. D {\bf 51}, 3436 (1995).
\bibitem{lpm2}R. Baier, Yu.~L.~Dokshitser, S.~Peigne, D.~Schiff, 
Phys. Lett. {\bf B345}, 277 (1995).
 
\end{references}
\end{document}